\documentclass[10pt,aps,prx,twocolumn,notitlepage,showpacs,superscriptaddress]{revtex4-1}
\usepackage{amsthm,amsmath,amsfonts,amssymb,verbatim,color}
\usepackage{graphicx}
\usepackage{bm}
\usepackage{epsfig,slashed}
\usepackage{amssymb}
\usepackage{amsfonts}
\usepackage{tikz}
\usepackage{xparse}
\usepackage{ifthen}
\usepackage{lipsum}
\usepackage[colorlinks=true,citecolor=blue,
linkcolor=blue,urlcolor=blue]{hyperref}
\usepackage[caption=false]{subfig}
\begin{document}

\newcommand{\tr}{\mathop{\mathrm{Tr}}}
\newcommand{\bsigma}{\boldsymbol{\sigma}}
\newcommand{\re}{\mathop{\mathrm{Re}}}
\newcommand{\im}{\mathop{\mathrm{Im}}}
\renewcommand{\b}[1]{{\boldsymbol{#1}}}
\newcommand{\diag}{\mathrm{diag}}
\newcommand{\sign}{\mathrm{sign}}
\newcommand{\sgn}{\mathop{\mathrm{sgn}}}
\renewcommand{\c}[1]{\mathcal{#1}}

\newcommand{\mb}{\bm}
\newcommand{\ua}{\uparrow}
\newcommand{\da}{\downarrow}
\newcommand{\ra}{\rightarrow}
\newcommand{\la}{\leftarrow}
\newcommand{\mc}{\mathcal}
\newcommand{\bs}{\boldsymbol}
\newcommand{\lra}{\leftrightarrow}
\newcommand{\nn}{\nonumber}
\newcommand{\half}{{\textstyle{\frac{1}{2}}}}
\newcommand{\mf}{\mathfrak}
\newcommand{\MF}{\text{MF}}
\newcommand{\IR}{\text{IR}}
\newcommand{\UV}{\text{UV}}
\newcommand{\sech}{\mathrm{sech}}

\title{Nematic order on the surface of a three-dimensional topological insulator}
\author{Rex Lundgren}
\affiliation{Department of Physics, The University of Texas at Austin, Austin, Texas 78712, USA}
\affiliation{Joint Quantum Institute, NIST/The University of Maryland, College Park, Maryland 20742, USA}
\author{Hennadii Yerzhakov}
\affiliation{Department of Physics, University of Alberta, Edmonton, Alberta T6G 2E1, Canada}
\author{Joseph Maciejko}
\affiliation{Department of Physics, University of Alberta, Edmonton, Alberta T6G 2E1, Canada}
\affiliation{Theoretical Physics Institute, University of Alberta, Edmonton, Alberta T6G 2E1, Canada}
\affiliation{Canadian Institute for Advanced Research, Toronto, Ontario M5G 1Z8, Canada}

\date\today

\begin{abstract}
We study the spontaneous breaking of rotational symmetry in the helical surface state of three-dimensional topological insulators due to strong electron-electron interactions, focusing on time-reversal invariant nematic order. Owing to the strongly spin-orbit coupled nature of the surface state, the nematic order parameter is linear in the electron momentum and necessarily involves the electron spin, in contrast with spin-degenerate nematic Fermi liquids. For a chemical potential at the Dirac point (zero doping), we find a first-order phase transition at zero temperature between isotropic and nematic Dirac semimetals. This extends to a thermal phase transition that changes from first to second order at a finite-temperature tricritical point. At finite doping, we find a transition between isotropic and nematic helical Fermi liquids that is second order even at zero temperature. Focusing on finite doping, we discuss various observable consequences of nematic order, such as anisotropies in transport and the spin susceptibility, the partial breakdown of spin-momentum locking, collective modes and induced spin fluctuations, and non-Fermi liquid behavior at the quantum critical point and in the nematic phase.
\end{abstract}

\maketitle


\section{Introduction}
Rotationally invariant Fermi liquids can spontaneously develop spatial anisotropy as a result of strong electron-electron interactions, a possibility first considered by Pomeranchuk~\cite{pomeranchuk1958}. In the simplest scenario, for sufficiently strong attractive interactions in the $l=2$ angular momentum channel the ground state energy of the Fermi liquid is lowered by a spontaneous quadrupolar distortion of the Fermi surface, leading to transport anisotropies and non-Fermi liquid behavior~\cite{oganesyan2001}. Alternatively, the resulting time-reversal and translationally invariant form of order, nematic order, can arise via thermal or quantum melting of translational symmetry-breaking stripe/smectic orders~\cite{2010ARCMP...1..153F}. There is strong experimental evidence for the existence of a nematic phase in quantum Hall states~\cite{PhysRevLett.82.394,Du1999389,PhysRevLett.83.820,PhysRevLett.83.824,PhysRevB.65.241313,PhysRevLett.105.176807,2011NatPh...7..845X,2016NatPh..12..191S},   high-temperature superconductors~\cite{2009AdPhy..58..699V, fernandes2014,ianfisherscience}, and Sr$_3$Ru$_2$O$_7$~\cite{2007Sci...315..214B,PhysRevB.81.081105}. On the theory side, nematic order has been studied in a wide variety of systems including quantum Hall states~\cite{PhysRevB.59.8065,PhysRevLett.84.1982,PhysRevB.64.115312,0953-8984-14-14-303,PhysRevLett.88.216802,PhysRevB.75.195433,PhysRevB.82.085102,PhysRevB.84.195124,PhysRevB.88.125137,PhysRevB.88.235124,PhysRevX.4.041050,PhysRevB.93.205401,2016arXiv160702178R}, graphene~\cite{PhysRevB.80.075108,1367-2630-10-11-113009}, two- and three-dimensional systems with quadratic band crossing~\cite{PhysRevLett.103.046811,PhysRevB.92.045117}, three-dimensional Dirac semimetals \cite{PhysRevB.93.041108}, dipolar Fermi gases~\cite{PhysRevLett.103.205301,2009NJPh...11j3003F,PhysRevA.81.023602}, high-temperature superconductors~\cite{PhysRevB.69.144505} and doped Mott insulators~\cite{1998Natur.393..550K}.

The surface of 3D topological insulators offers a new type of gapless matter, the 2D helical Dirac fermion, which differs qualitatively from conventional Fermi systems due to the phenomenon of spin-momentum locking~\cite{RevModPhys.82.3045,RevModPhys.83.1057}. This begs the question whether criteria for electronic instabilities and the nature of possible broken-symmetry states on the surface of a 3D topological insulator differ from those of conventional 2D Fermi systems. While previous work has focused largely on superconducting~\cite{santos2010,roy2013,nandkishore2013,dassarma2013,grover2014,ponte2014,neupert2015,witczak-krempa2016,zerf2016,jian2016} and time-reversal breaking~\cite{xu2010a,xu2010,kim2011,jiang2011,ghaemi2012,nogueira2013,sitte2013,bahri2014,mendler2015,neupert2015,PhysRevB.94.115429} instabilities, little attention has been devoted to nematic instabilities, with the exception of Ref.~\cite{roy2015} which studies the spontaneous breaking of a discrete rotation symmetry on the surface of a topological Kondo insulator with multiple Dirac cones.

Our focus here is the isotropic-nematic phase transition on the surface of a 3D topological insulator with a single rotationally invariant Dirac cone. For an undoped system (chemical potential at the Dirac point) one always has continuous rotational invariance in the low-energy limit; for a doped system our theory could apply to a number of experimentally realized topological insulators with very nearly circular Fermi surfaces, such as Bi$_2$Se$_3$~\cite{xia2009,*hsieh2009,*pan2011}, Bi$_2$Te$_2$Se~\cite{neupane2012,*neupane2013}, Sb$_x$Bi$_{2-x}$Se$_2$Te~\cite{neupane2012}, Bi$_{1.5}$Sb$_{0.5}$Te$_{1.7}$Se$_{1.3}$~\cite{kim2013}, Tl$_{1-x}$Bi$_{1+x}$Se$_{2-\delta}$~\cite{kuroda2013}, strained $\alpha$-Sn on InSb(001)~\cite{barfuss2013}, and strained HgTe~\cite{crauste2013}. A phenomenological Landau Fermi liquid theory of the topological surface state developed earlier by two of us showed that an isotropic-nematic quantum phase transition can occur in the doped system for a sufficiently negative value of the $l=2$ ``projected'' Landau parameter $\bar{f}_2$~\cite{PhysRevLett.115.066401}, in full analogy with the standard Pomeranchuk instability. In this work we construct a field theory of the transition, investigate both the doped and undoped limits, and extend our analysis to nonzero temperatures. In the doped limit we find a continuous transition already at zero temperature, with a breakdown of helical Fermi liquid behavior at the quantum critical point and in the nematic phase, in analogy with the spin-degenerate problem~\cite{oganesyan2001}. The nematic phase exhibits a partial breakdown of spin-momentum locking, in the sense that spin and momentum are no longer orthogonal to each other except at certain discrete points on the Fermi surface. Other unusual observable consequences of the spin-orbit coupled nature of nematic order in this system include anisotropy in the in-plane spin susceptibility in the absence of time-reversal symmetry breaking and the generation of spin fluctuations from nematic fluctuations at finite frequency. At zero doping the isotropic-nematic transition is first-order at zero temperature and becomes continuous at a finite-temperature tricritical point.

The paper is organized as follows. In Sec.~\ref{sec:Model}, we introduce our model and argue that strong spin-orbit coupling on the topological insulator surface warrants a novel type of nematic order parameter that mixes charge and spin degrees of freedom. In Sec.~\ref{sec:Mean}, we construct a mean-field theory of the isotropic-nematic transition at both zero and finite temperature and discuss the consequences of nematic order for electronic properties at the mean-field level. Sec.~\ref{sec:fluctuations} discusses fluctuation effects beyond mean-field theory, namely, collective modes and their effect on electronic properties. A brief conclusion is given in Sec.~\ref{sec:conclusions}.

\section{Model and Nematic Order Parameter}\label{sec:Model}

In this section, we introduce our field-theoretic model for the isotropic-nematic transition on the surface of a 3D topological insulator. We follow largely the approach of Ref.~\cite{oganesyan2001}, with important caveats due to the presence of strong spin-orbit coupling, as will be seen below. While nematic order in 2D electron gases with Rashba spin-orbit coupling has been studied before~\cite{berg2012,ruhman2014}, such systems have two degenerate concentric Fermi surfaces and are thus qualitatively distinct from the single, nondegenerate helical Fermi surface considered here.

The Hamiltonian that describes the noninteracting gapless surface state of a topological insulator with a single Dirac cone is given by \cite{RevModPhys.82.3045,RevModPhys.83.1057} (in units where $\hbar=k_B=1$)
\begin{align}\label{H0}
H_0=\int\frac{d^2k}{(2\pi)^2}\psi_\b{k}^\dag (h(\b{k})-\mu)\psi_\b{k},
\end{align}
where $\psi_\b{k}=(\psi_{\b{k}\ua},\psi_{\b{k}\da})$ is a two-component Dirac spinor, $v_F$ is the Fermi velocity, $\mu$ is the chemical potential, and
\begin{align}
h(\b{k})=v_F\hat{\b{z}}\cdot(\b{\sigma}\times\b{k})=v_F\left(
\begin{array}{cc}
0 & ike^{-i\theta_\b{k}} \\
-ike^{i\theta_\b{k}} & 0
\end{array}\right),
\end{align}
where $\b{\sigma}$ is a vector of Pauli matrices, $\theta_\b{k}=\tan^{-1}(k_y/k_x)$ and $k=\sqrt{k_x^2+k_y^2}$. The Hamiltonian \eqref{H0} has a continuous spatial $SO(2)$ rotation symmetry, $[J_z,h(\b{k})-\mu]=0$, where
\begin{align}\label{Jz}
J_z=-i\frac{\partial}{\partial\theta_\b{k}}+\frac{1}{2}\sigma_z,
\end{align}
is the $z$ component of total angular momentum.

In order to study the isotropic-nematic transition we need a suitable microscopic definition of the nematic order parameter in terms of the fermionic fields $\psi,\psi^\dag$. In general, nematic order is described by a quadrupolar order parameter $Q_{ab}$ which transforms as a real, traceless symmetric rank-two tensor under rotations~\cite{p1995physics}. Because of spin-orbit coupling, here the relevant rotations are simultaneous rotations in real space and spin space, generated by the total angular momentum (\ref{Jz}). Therefore, unlike for spin rotationally invariant Fermi liquids~\cite{oganesyan2001} the nematic order parameter can involve both the spatial (charge) and spin degrees of freedom of the electron. To lowest order in the electron momentum, the appropriate generalization of the nematic order parameter considered in Ref.~\cite{oganesyan2001} for spin rotationally invariant Fermi liquids to the surface state of 3D topological insulators is
\begin{align}\label{OP}
\hat{Q}_{ab}(\b{r})=-\frac{i}{k_A}\psi^\dag(\b{r})(\sigma_a\overset{\leftrightarrow}{\partial_b}+\sigma_b\overset{\leftrightarrow}{\partial_a}-\delta_{ab}\b{\sigma}\cdot\overset{\leftrightarrow}{\b{\partial}})\psi(\b{r}),
\end{align}
where $a,b=1,2$, and $\overset{\leftrightarrow}{\b{\partial}}=(\overset{\leftrightarrow}{\partial_x},\overset{\leftrightarrow}{\partial_y})$ is a vector of symmetrized derivatives whose action is defined as $\psi^\dag\overset{\leftrightarrow}{\partial_a}\psi\equiv\frac{1}{2}(\psi^\dag\partial_a\psi+(\partial_a\psi^\dag)\psi)$. This ensures $\hat{Q}_{ab}(\b{r})$ is a Hermitian operator. Finally, the parameter $k_A$ is defined differently depending on whether one is in the doped or undoped limit. We consider that four-fermion interactions, to be written out explicitly below, only act within a high-energy cutoff that can be converted to a momentum cutoff $\Lambda$ by dividing by $v_F$. In the undoped limit $\mu=0$, we define $k_A\equiv\Lambda$ and the order parameter is local in space. This order parameter was first introduced by one of us in the context of nematic instabilities of the Majorana surface state of superfluid $^3$He-$B$~\cite{park2015}, and its 3D analog was proposed as an order parameter for parity-breaking phases of spin-orbit coupled bulk metals~\cite{fu2015,norman2015}. In the doped limit, defined as $\mu\gg v_F\Lambda$, only (angular) degrees of freedom on the Fermi surface are relevant and we define $k_A\equiv|\b{\partial}|$~\cite{PhysRevB.75.115103}.

In the spirit of Ref.~\cite{oganesyan2001}, we consider an attractive four-fermion interaction in the quadrupolar ($l=2$) channel,
\begin{align}\label{Hint}
H_\textrm{int}=-\frac{f_2}{4}\int d^2r\tr\left(\hat{Q}(\b{r})^2\right),
\end{align}
where $\tr$ denotes a trace over the spatial (nematic) indices $a,b$. The action in imaginary time is then
\begin{align}
S[\psi^\dag,\psi]=&\int_0^{1/T}d\tau\int d^2r\bigg[\psi^\dag(\partial_\tau-iv_F\hat{\b{z}}\cdot(\b{\sigma}\times\b{\partial})-\mu)\psi\nonumber \\
&-\frac{f_2}{4}\tr\left(\hat{Q}(\b{r})^2\right)\bigg].
\label{Main_Action}
\end{align}
As our focus is the vicinity of the isotropic-nematic transition, interactions in other angular momentum channels have been ignored. Indeed, in the doped limit, as long as such interactions are less than the critical value for a $l\neq 2$ Pomeranchuk instability, they will simply lead to a finite renormalization of physical quantities such as the Fermi velocity~\cite{PhysRevLett.115.066401}. While the phenomenological Landau Fermi liquid description does not strictly apply to the undoped case, we will assume in this case that interactions in $l\neq 2$ channels are sufficiently weak so there are no competing instabilities.

\section{Mean-Field Theory}\label{sec:Mean}

To investigate a possible isotropic-nematic phase transition in the action (\ref{Main_Action}), we analyze it in the mean-field approximation. Introducing a real auxiliary scalar field $Q_{ab}(\tau,\b{r})$ to decouple the four-fermion term via the Hubbard-Stratonovich transformation, we have
\begin{widetext}
\begin{align}
S[\psi^\dag,\psi,Q_{ab}]=\int_0^{1/T} d\tau \int d^2r\bigg[\psi^\dag(\partial_\tau-iv_F\hat{\b{z}}\cdot(\b{\sigma}\times\b{\partial})-\mu)\psi 
-\frac{iQ_{ab}}{k_A}\psi^\dag(\sigma_a\overset{\leftrightarrow}{\partial_b}+\sigma_b\overset{\leftrightarrow}{\partial_a}-\delta_{ab}\b{\sigma}\cdot\overset{\leftrightarrow}{\b{\partial}})\psi+\frac{1}{f_2}\tr(Q^2)\bigg].
\label{eff_action_mean}
\end{align}
\end{widetext}
Assuming a uniform and static order parameter $Q_{ab}(\tau,\b{r})=\bar{Q}_{ab}$, and integrating out the fermions, we obtain the following saddle-point free energy density,
\begin{equation}
\c{F}(\bar{Q})=\frac{2}{f_2}\bar{Q}^2-\frac{T}{V}\sum_{ik_n}\sum_{\b{k}}\ln\left[(k_n-i\mu)^2+\epsilon_\b{k}(\bar{Q})^2\right],
\end{equation}
where $k_n=(2n+1)\pi T$, $n\in\mathbb{Z}$ is a fermionic Matsubara frequency. We have rotated the order parameter such that $\bar{Q}_{11}=-\bar{Q}_{22}=0$, $\bar{Q}_{12}=\bar{Q}_{21}=\bar{Q}$ without loss of generality (corresponding to the principal axes of the distorted Fermi surface being parallel to the $x$ and $y$ axes~\footnote{There is a $\pi/4$ angle difference between the naive orientation of $\hat{Q}_{ab}$ in Eq.~(\ref{OP}) and the principal axes of the distorted Fermi surface, or equivalently, the orientation of the effective spinless nematic order parameter that results from projection to the Fermi surface~\cite{PhysRevLett.115.066401}.}), and
\begin{equation}\label{MFdispersion}
\epsilon_\b{k}(\bar{Q})=\sqrt{(\epsilon_{\b{k}}^0)^2-4\bar{Q}\epsilon_{\b{k}}^0\frac{k}{k_A} \cos2\theta_{\b{k}}+4\bar{Q}^2\bigg(\frac{k}{k_A}\bigg)^2},
\end{equation}
is the mean-field dispersion relation of fermionic quasiparticles in the nematic phase (for $\bar{Q}\neq 0$), where $\epsilon_{\b{k}}^0=v_Fk$ is the dispersion relation in the isotropic phase. This corresponds to an anisotropic Dirac cone (in the doped limit, $\epsilon_\b{k}(\bar{Q})$ is only meant to model the dispersion of quasiparticles on the Fermi surface, with $k\approx k_F\equiv\mu/v_F$). Here $k_A$ is to be understood in momentum space, i.e., $k_A=\Lambda$ in the undoped limit and $k_A=k$ in the doped limit. Performing the sum over Matsubara frequencies, and ignoring constant terms, we obtain
\begin{align}\label{free_energy_density}
\c{F}(\bar{Q})=\frac{2}{f_2}\bar{Q}^2-T\sum_s\int\frac{d^2k}{(2\pi)^2}
\ln\left(1+e^{-(s\epsilon_\b{k}(\bar{Q})-\mu)/T}\right),
\end{align}
where $s=\pm 1$ corresponds to the upper and lower branches of the Dirac cone, respectively, and we have taken the infinite volume limit $V\rightarrow\infty$. At zero temperature, Eq.~(\ref{free_energy_density}) becomes the ground state energy density,
\begin{align}\label{ground_energy_density}
\c{E}(\bar{Q})=\frac{2}{f_2}\bar{Q}^2-\frac{1}{2}\sum_s\int\frac{d^2k}{(2\pi)^2}|s\epsilon_\b{k}(\bar{Q})-\mu|.
\end{align}
In the following our analysis is performed at constant $\mu$.

\subsection{Undoped limit}

We first evaluate the free energy density in the undoped limit ($\mu=0$). At zero temperature, we have
\begin{align}\label{ground_energy_density_mu0}
\c{E}(\bar{Q})=\frac{2}{f_2}\bar{Q}^2-\int_{|\b{k}|<\Lambda}\frac{d^2k}{(2\pi)^2}\epsilon_\b{k}(\bar{Q}),
\end{align}
where we have imposed the momentum cutoff $\Lambda$. The integral over momentum can be performed exactly, and we obtain
\begin{align}\label{FreeEnergyMu0}
\c{E}(\Delta)=\frac{v_F\Lambda^3}{3\pi^2}\left[\frac{\Delta^2}{\lambda}-|\Delta-1|E\left(-\frac{4\Delta}{(\Delta-1)^2}\right)\right],
\end{align}
where $E(m)$ is the complete elliptic integral of the second kind, and we define a dimensionless nematic order parameter $\Delta=2\bar{Q}/v_F\Lambda$ and a dimensionless interaction strength $\lambda=2f_2\Lambda/3\pi^2v_F$. A strongly first-order isotropic-nematic transition is found at a critical value $\lambda_c\approx 2.13$, with a jump of order one in the order parameter $\Delta$ at the transition, corresponding to a value of $\bar{Q}$ on the order of the high-energy cutoff $v_F\Lambda$. This is to be expected since $\bar{Q}$ has units of energy, and in the undoped limit the only energy scale in the problem is the cutoff (the critical value of the interaction strength $f_2$ is also determined by the cutoff, since the interaction (\ref{Hint}) is perturbatively irrelevant at the Dirac point). Expanding (\ref{FreeEnergyMu0}) in powers of $\Delta$ in the limit $|\Delta|\ll 1$, we find
\begin{align}\label{GEDexpansion}
\c{E}(\Delta)-\c{E}(0)=\frac{v_F\Lambda^2}{3\pi^2}\left[\left(\frac{1}{\lambda}-\frac{\pi}{8}\right)\Delta^2+\ldots\right],
\end{align}
hence the limit of metastability of the isotropic phase (corresponding to the divergence of the nematic susceptibility) is $\lambda^*=8/\pi\approx 2.55$, but this is preempted by the first-order transition at $\lambda_c\approx 2.13$. The limit of metastability of the nematic phase can be found numerically, and is $\lambda^{**}\approx 1.90$.

\begin{figure}[t]
\includegraphics[width=1.0\linewidth]{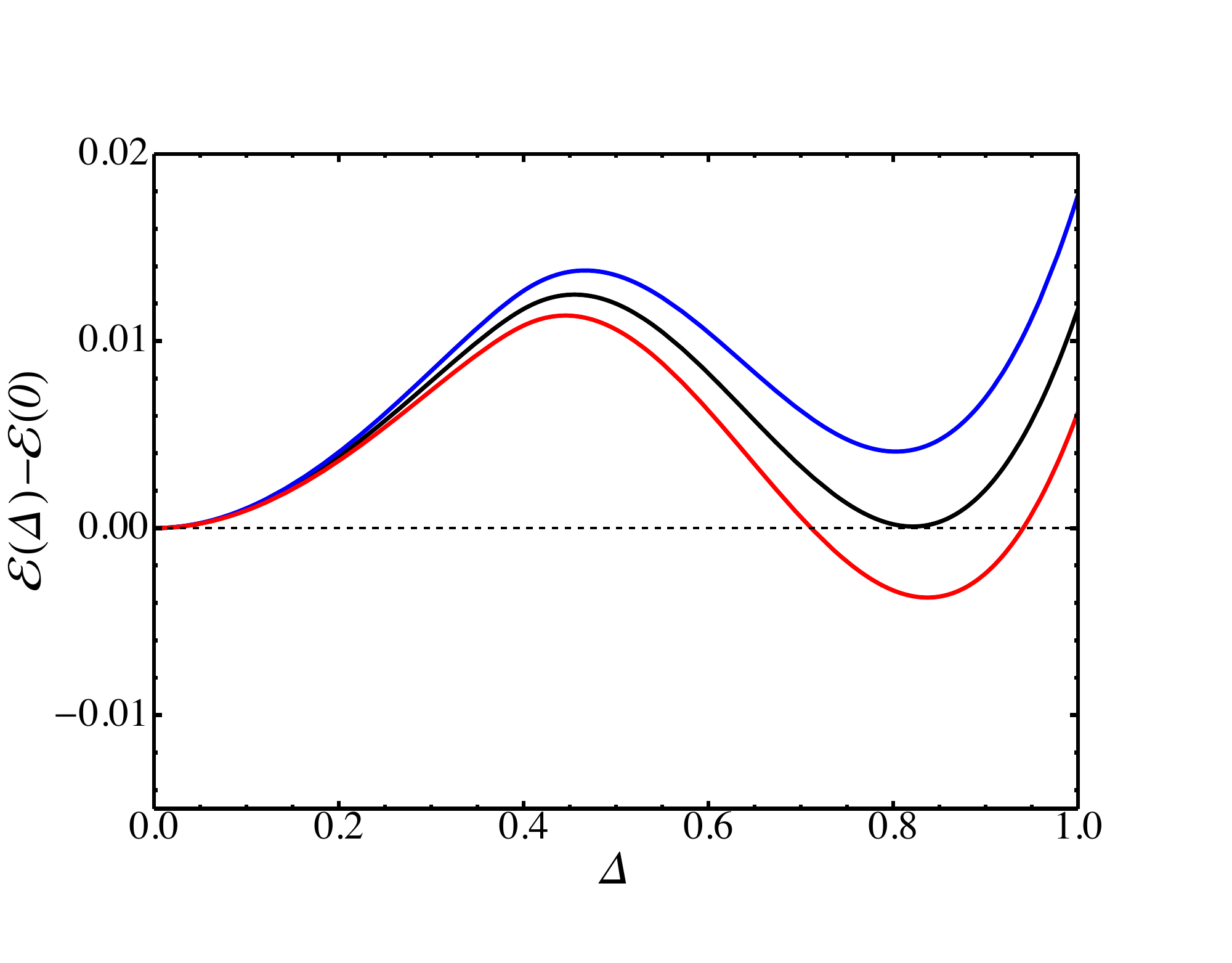}
\caption{First-order isotropic-nematic quantum phase transition in the undoped limit ($\mu=0$). Plots of the mean-field ground state energy density $\c{E}(\Delta)$ in units of $v_F\Lambda^3/3\pi^2$ are given as a function of the dimensionless nematic order parameter $\Delta$, for $\lambda<\lambda_c$ (blue curve), $\lambda=\lambda_c$ (black curve), and $\lambda>\lambda_c$ (red curve), where $\lambda$ is the dimensionless interaction strength with critical value $\lambda_c\approx 1.31$ at the transition. The leading correction to linear dispersion is given by $\alpha=-0.61$.}
\label{fig:transition}
\end{figure}

\begin{figure*}[ht!]
\subfloat[][]{
 \includegraphics[width=.45\linewidth]{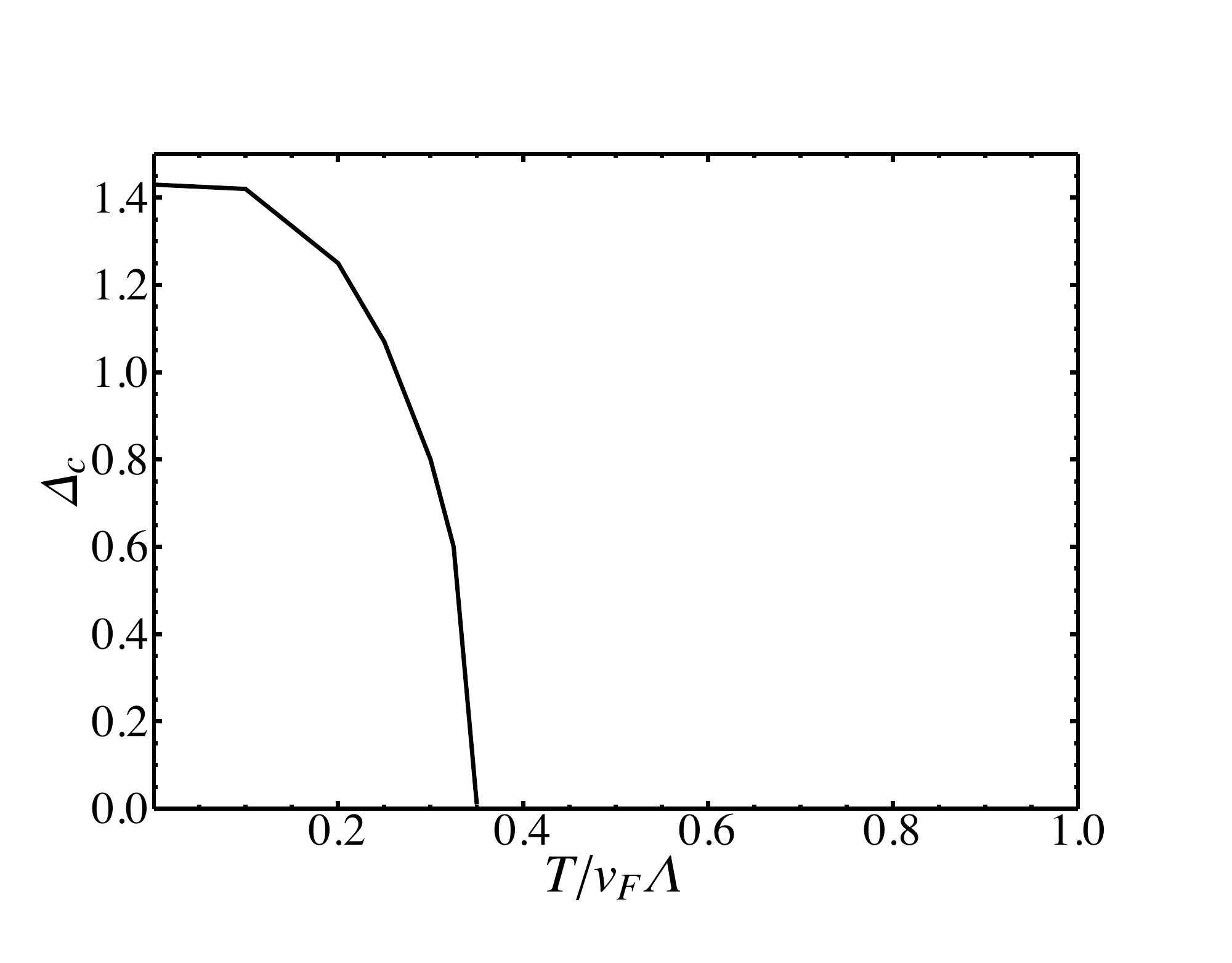}
 \label{fig:deltacrit}
}
\subfloat[][]{
 \includegraphics[width=.45\linewidth]{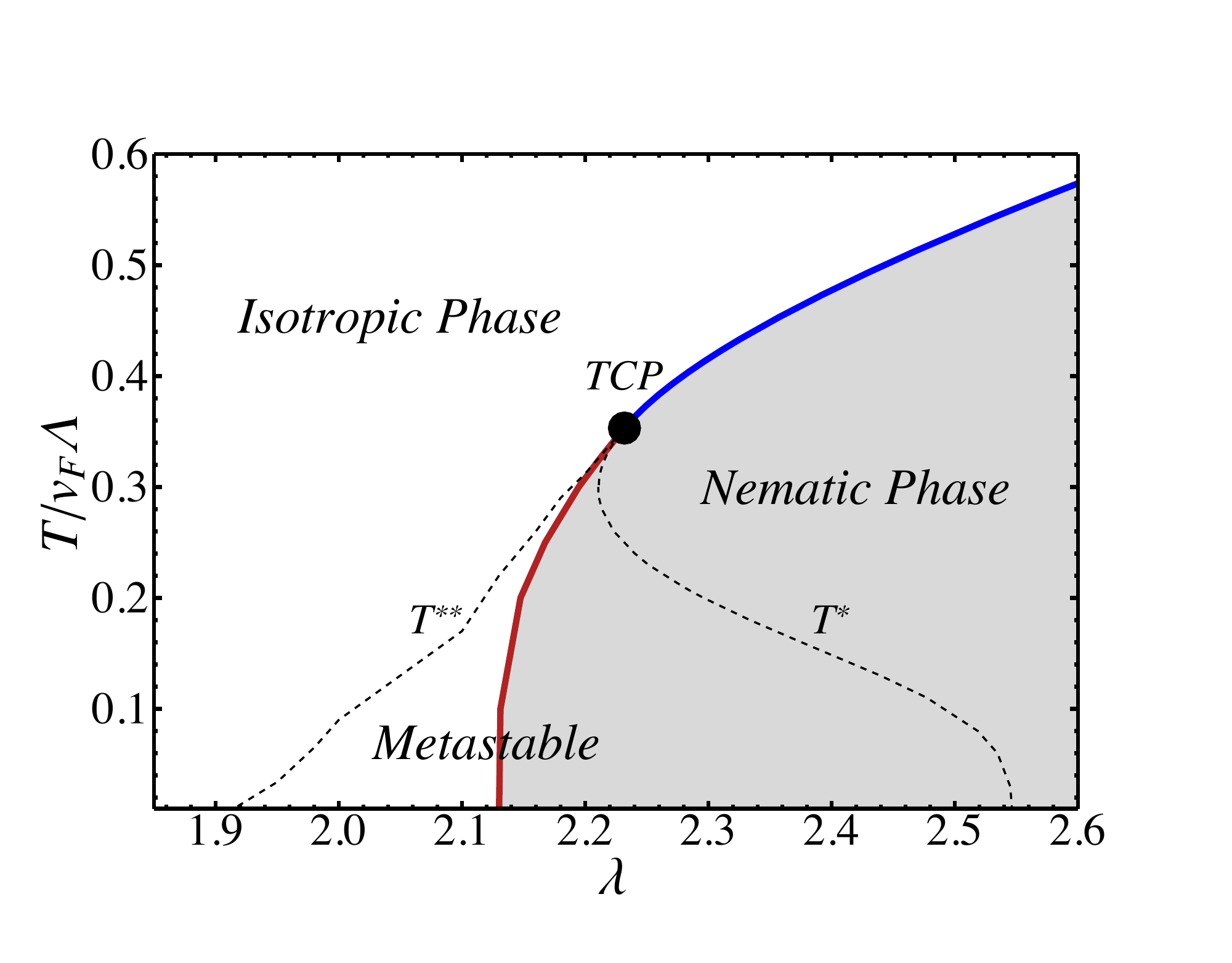}
 \label{fig:phase_dia}
}
\caption{(color online) Finite-temperature isotropic-nematic transition in the undoped limit ($\mu=0$): (a) Jump in dimensionless nematic order parameter at the first-order phase transition as a function of temperature; (b) Mean-field phase diagram in the plane of of temperature $T$ and dimensionless interaction strength $\lambda$. A first-order transition (red line) at low temperature turns into a continuous transition (blue line) above a tricritical point (black dot). Dotted lines correspond to limits of metastability of the isotropic ($T^*$) and nematic ($T^{**}$) phases.}
\end{figure*}

The magnitude of the order parameter jump at the transition can be reduced somewhat by considering the effects of nonzero band curvature at the Dirac point, i.e., deviations from a perfectly linear dispersion (which are present anyway in real topological insulator materials). In other words, we replace $v_F$ in the noninteracting dispersion $\epsilon_\b{k}^0$ by a $k$-dependent Fermi velocity
\begin{align}\label{ModifFermiVelocity}
v_F(k)=v_F\left[1+\alpha\left(\frac{k}{\Lambda}\right)^2+\ldots\right],
\end{align}
with the dimensionless parameter $\alpha$ representing the leading correction. Such corrections are formally irrelevant in the low-energy limit $k\ll\Lambda$ but affect the free energy~\cite{oganesyan2001,PhysRevX.4.041050}, which depends on the noninteracting dispersion at all wavevectors up to the cutoff. In the presence of such terms the energy density cannot be evaluated analytically and one must resort to numerical integration. A typical plot of the ground state energy density in the vicinity of the transition for nonzero $\alpha$ is given in Fig.~\ref{fig:transition}. We have found that negative values of $\alpha$ reduce both the critical interaction strength and order parameter jump at the transition below their values for a strictly linear dispersion.

The apperance of a first-order transition is somewhat surprising, since Landau theory predicts a continuous isotropic-nematic transition in 2D (unlike in 3D, there are no cubic invariants). Expanding the quasiparticle dispersion relation $\epsilon_\b{k}(\bar{Q})$ in powers of $\bar{Q}$ in Eq.~(\ref{ground_energy_density_mu0}), and performing the integral over $\b{k}$, we obtain the Landau theory
\begin{align}\label{LandauExpansion}
\c{E}(\Delta)-\c{E}(0)\stackrel{?}{=}\frac{v_F\Lambda^2}{3\pi^2}\left[\left(\frac{1}{\lambda}-\frac{\pi}{8}\right)\Delta^2+\sum_{n=2}^\infty c_{2n}\Delta^{2n}\right],
\end{align}
where $c_{2n}<0$ for all $n\geq 2$. We have checked that the only way to get a quartic term $\propto\Delta^4$ with positive coefficient is to consider a $k$-dependent Fermi velocity $v_F(k)$ that becomes negative at a certain value of $k$ below the cutoff $\Lambda$, in clear contradiction with the assumption of a single Dirac point in the low-energy spectrum. Therefore, the Landau theory (\ref{LandauExpansion}) is unbounded from below for sufficiently large $\Delta$, in disagreement with the exact energy density (\ref{FreeEnergyMu0}) which behaves qualitatively like in Fig.~\ref{fig:transition}. As a result, there must be nonanalytic terms in Eq.~(\ref{FreeEnergyMu0}), but missed by the Landau expansion around $\Delta=0$, that stabilize the energy density. Such nonanalytic terms are ultimately responsible for the first-order character of the phase transition. In fact, for $|\Delta|\gg 1$ the energy density (\ref{FreeEnergyMu0}) becomes
\begin{align}
\c{E}(\Delta)-\c{E}(0)\approx\frac{v_F\Lambda^2}{3\pi^2}\left(\frac{\Delta^2}{\lambda}-\frac{\pi}{2}|\Delta|\right),\hspace{5mm}|\Delta|\gg 1.
\end{align}
Thus the energy density is stabilized at large $\Delta$ by the ``bare'' (tree-level) mass term $\Delta^2/\lambda$, which grows faster than the negative $|\Delta|$ term coming from the one-loop fermion determinant, i.e., the integral over quasiparticle energies in Eq.~(\ref{ground_energy_density_mu0}). The latter is in fact negative for all $\Delta$. We note that a first-order Ising nematic transition at zero temperature was also found for a model of interacting electrons on the square lattice~\cite{PhysRevB.70.155110}. In this case van Hove singularities in the quasiparticle density of states, corresponding to Lifshitz transitions tuned by the value of $\bar{Q}$, are responsible for nonanalyticities in the energy density and the first-order character of the transition.

At finite temperature the free energy density in the undoped limit is given by
\begin{align}
\c{F}(\bar{Q})=\frac{2}{f_2}\bar{Q}^2-T\sum_s\int_{|\b{k}|<\Lambda}\frac{d^2k}{(2\pi)^2}\ln\left(1+e^{-s\epsilon_\b{k}(\bar{Q})/T}\right).
\end{align}
In the remainder of this section we focus on the limit of strict linear dispersion $v_F(k)=v_F$. The integral over the magnitude of $k$ can be evaluated analytically in terms of dilogarithms $\mathrm{Li}_2(x)$ and trilogarithms $\mathrm{Li}_3(x)$; the remaining angular integral must be performed numerically. In Fig.~\ref{fig:deltacrit} we plot the jump $\Delta_c$ in the order parameter at the transition as a function of temperature $T$. The jump decreases smoothly from its value at zero temperature, eventually vanishing above a certain temperature $T_\textrm{TCP}$ corresponding to a tricritical point; for $T>T_\textrm{TCP}$ the transition is continuous (a similar behavior was found in Ref.~\cite{PhysRevB.70.155110}). Since $\Delta$ vanishes at the tricritical point, to find $T_\textrm{TCP}$ we expand the free energy density (\ref{free_energy_density}) in powers of $\Delta$. To describe the tricritical point we must expand to sixth order,
\begin{align}\label{LandauFiniteT}
\c{F}(\Delta,T)-\c{F}(0,T)=\frac{v_F\Lambda^3}{3\pi^2}\left(a_2\Delta^2+a_4\Delta^4+a_6\Delta^6\right),
\end{align}
where $a_2,a_4,a_6$ are functions of $T$. We find that $a_6>0$ for $0.2\lesssim T/v_F\Lambda\lesssim 0.6$, which comprises the tricritical point (Fig.~\ref{fig:deltacrit}). The tricritical point $(T_\textrm{TCP},\lambda_\textrm{TCP})$ is found from the condition $a_2=a_4=0$, from which we find $T_\textrm{TCP}/v_F\Lambda\approx 0.35$ and $\lambda_\textrm{TCP}\approx 2.23$. The finite-temperature phase diagram is shown in Fig.~\ref{fig:phase_dia}, in which we also plot the limits of metastability of the isotropic ($T^*$) and nematic ($T^{**}$) phases. Note that the first-order phase boundary and limits of metastability are obtained from the numerically evaluated, exact free energy density (\ref{free_energy_density}) rather than from the Landau expansion (\ref{LandauFiniteT}), which is accurate only in the vicinity of the continuous transition. Strictly speaking, the finite-temperature phase transition for $T>T_\textrm{TCP}$ is a Kosterlitz-Thouless transition and the nematic phase only exhibits quasi-long-range order at finite $T$ (but is truly long-range ordered at $T=0$).

At the mean-field level, the nematic phase is a theory of noninteracting Dirac quasiparticles with anisotropic dispersion, with Hamiltonian $H_\text{MF}=\sum_{\b{k}}\psi^{\dagger}_{\b{k}}\c{H}_\b{k}^{\phantom{\dagger}}\psi_{\b{k}}^{\phantom{\dagger}}$ where
\begin{align}
\c{H}_\b{k}=v_F\hat{\b{z}}\cdot(\b{\sigma}\times\b{k})+\frac{\bar{Q}_{ab}}{\Lambda}(\sigma_ak_b+\sigma_bk_a-\delta_{ab}\b{\sigma}\cdot\b{k}).
\end{align}
Without loss of generality we choose $\bar{Q}_{12}=\bar{Q}_{21}=\bar{Q}$, $\bar{Q}_{11}=-\bar{Q}_{22}=0$, and thus
\begin{align}
\c{H}_\b{k}=v_F\hat{\b{z}}\cdot(\b{\sigma}\times\b{k})+\frac{2\bar{Q}}{\Lambda}(\sigma_xk_y+\sigma_yk_x).
\end{align}
The velocities in the $x$ and $y$ directions (i.e., parallel to the principal axes of the nematic order parameter) at the Dirac point are
\begin{align}
v_x=v_F|1-\Delta|,\hspace{5mm}v_y=v_F|1+\Delta|.
\end{align}
Away from $\Delta=\pm 1$, the density of states remains linear near the Dirac point, $\mathcal{N}(\epsilon)\propto|\epsilon|$. In the limit of strict linear dispersion $v_F(k)=v_F$, the value $\Delta=1$ ($\Delta=-1$) thus corresponds to a Lifshitz transition where the quasiparticle dispersion vanishes along $x$ ($y$) and degenerates into the intersection of two planes, i.e., a quasi-1D Dirac dispersion with formally infinite density of states. In the presence of nonzero band curvature however [Eq.~(\ref{ModifFermiVelocity})], this degeneracy is lifted, and the flat direction acquires a cubic dispersion at small momenta,
\begin{align}
\epsilon_\b{k}(\Delta=1)\approx v_F\sqrt{4k_y^2+\frac{\alpha^2}{\Lambda^4}k_x^6},\,\b{k}\rightarrow 0,
\end{align}
with $k_x$ and $k_y$ interchanged for $\Delta=-1$. This corresponds to a density of states of the form $\mathcal{N}(\epsilon)\propto|\epsilon|^{1/3}$ near the Dirac point $\epsilon=0$.

\begin{figure}
\includegraphics[width=0.9\linewidth]{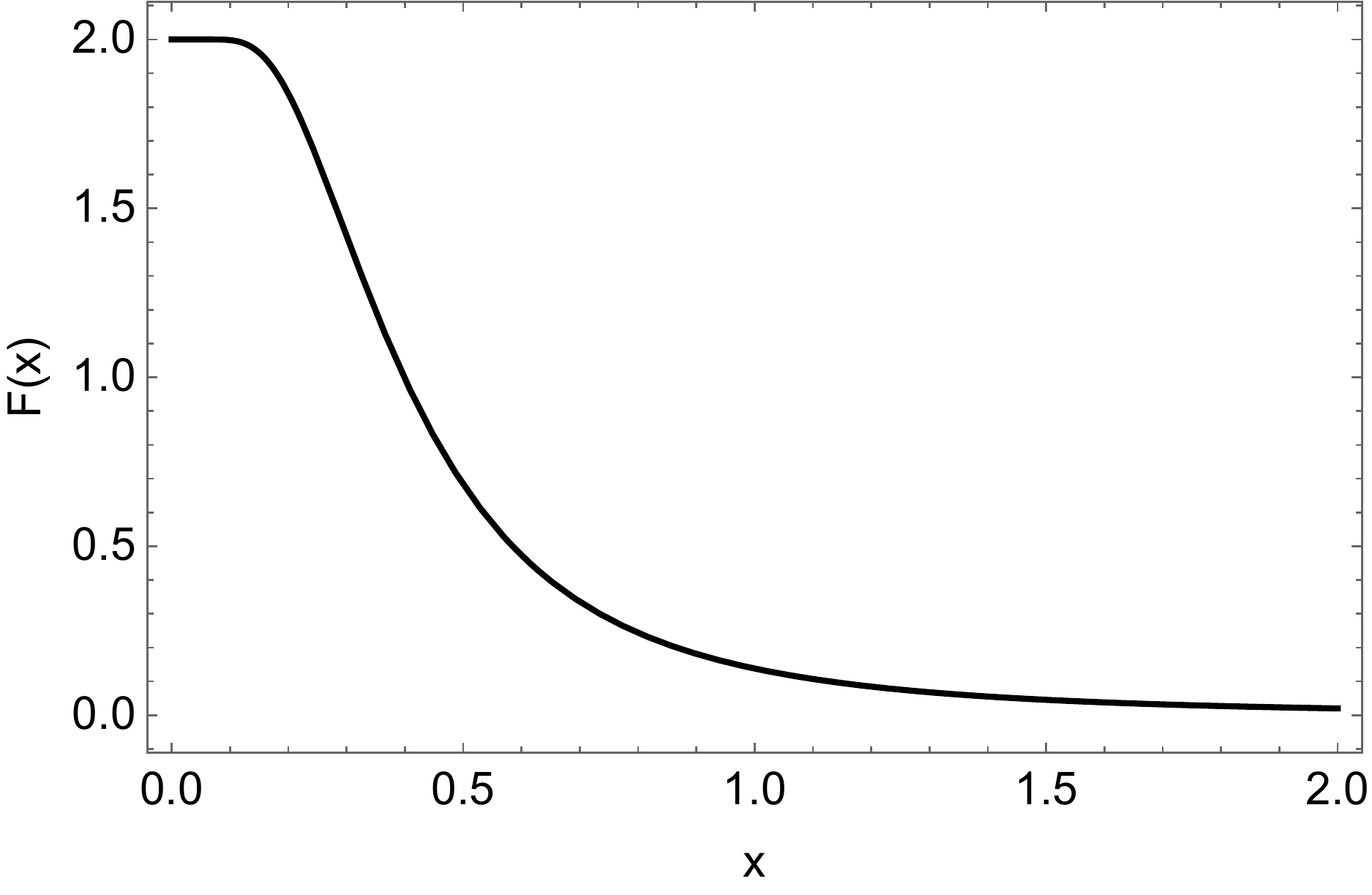}
\caption{Plot of the dimensionless function $F(x)$ defined in Eq.~(\ref{functionF}).}
\label{fig:F}
\end{figure}

An interesting signature of the unusual type of nematic order described here is anisotropy in the in-plane spin susceptibility in the absence of any time-reversal symmetry breaking. To compute the spin susceptibility we augment the mean-field Hamiltonian matrix (\ref{mean_electron}) with a Zeeman term,
\begin{align}
\delta\c{H}_\b{k}^Z=-\frac{1}{2}g\mu_B\b{B}\cdot\bsigma,
\end{align}
where $g$ is the $g$-factor, $\mu_B$ is the Bohr magneton, and $\b{B}$ is an in-plane magnetic field. To linear order in $\Delta$, we find
\begin{align}\label{DeltaChi}
\chi_{xx}(T)-\chi_{yy}(T)=\frac{g^2\mu_B^2\Lambda}{8\pi v_F}F\left(\frac{T}{v_F\Lambda}\right)\Delta(T),
\end{align}
where $\chi_{ij}(T)$ is the spin susceptibility tensor at temperature $T$, $\Delta(T)$ is the dimensionless nematic order parameter at temperature $T$, and $F$ is a smooth function of temperature (Fig.~\ref{fig:F}) defined as
\begin{align}\label{functionF}
F(x)=x\int_0^{1/x}dy\left[\sinh y+y\left(y\tanh\frac{y}{2}-1\right)\right]\sech^2\frac{y}{2}.
\end{align}
Thus anisotropy in the in-plane susceptibility is a direct measure of nematic order. For $T>T_\text{TCP}$, the transition is continuous (blue curve in Fig.~\ref{fig:phase_dia}) thus $\Delta(T)$ is small near $T_c$ and the expression (\ref{DeltaChi}) can be used in the vicinity of the transition. We thus expect
\begin{align}
\chi_{xx}(T)-\chi_{yy}(T)\propto F\left(\frac{T_c}{v_F\Lambda}\right)\Delta(T)\propto (T_c-T)^\beta,
\end{align}
on the nematic side of the transition, for $(T_c-T)/T_c\ll 1$. Thus the susceptibility anisotropy can give a direct measure of the order parameter critical exponent $\beta$, which is $1/2$ in mean-field theory. In the first-order region, since $\Delta$ may not be small Eq.~(\ref{DeltaChi}) cannot be directly used, but we nonetheless expect the anisotropy to be nonzero everywhere in the nematic phase and to vanish in the isotropic phase.

From a qualitative standpoint, the observation of in-plane spin susceptibility anisotropy in the absence of time-reversal symmetry breaking distinguishes the unusual type of nematic order discussed here from other types of order. For conventional nematic order in spin rotationally invariant systems~\cite{oganesyan2001}, the breaking of rotation symmetry is in the charge sector and does not cause anisotropy in the spin sector. In-plane ferromagnetic order would lead to anisotropy in the spin response, but requires time-reversal symmetry breaking.

\subsection{Doped limit}
\label{sec:MFDoped}

In the doped limit $\mu\gg v_F\Lambda$, the cutoff is imposed around the Fermi surface,
\begin{align}
\int_{|k-k_F|<\Lambda}\frac{d^2k}{(2\pi)^2}\equiv\int_{k_F-\Lambda}^{k_F+\Lambda}\frac{dk\,k}{2\pi}\int_0^{2\pi}\frac{d\theta_\b{k}}{2\pi},
\end{align}
where $k_F\equiv\mu/v_F$ is the (isotropic) Fermi momentum of noninteracting electrons. We obtain the ground state energy density (\ref{ground_energy_density}) to leading order in $\Lambda/k_F$ as
\begin{align}
\c{E}(\bar{Q})-\c{E}(0)=\left(\frac{2}{f_2}-\mathcal{N}(\mu)\right)\bar{Q}^2+\frac{\mathcal{N}(\mu)}{4\mu^2}\bar{Q}^4+\c{O}(\bar{Q}^6),
\label{S_eff_doped}
\end{align}
where $\mathcal{N}(\mu)=\mu/(2\pi v_F^2)$ is the noninteracting density of states at the Fermi surface. Since the coefficient of the $\bar{Q}^4$ term is positive, we therefore find a continuous quantum phase transition at a critical value of the interaction strength $f_2$ given by
\begin{align}\label{QCP}
\mathcal{N}(\mu)f_2=2.
\end{align}
From general considerations we expect a line of finite-$T$ Kosterlitz-Thouless phase transitions that terminates at this quantum critical point. We note also that Eq.~(\ref{QCP}) corresponds precisely to the $l=2$ Pomeranchuk criterion
\begin{align}
\bar{F}_2=-1,
\end{align}
derived from a phenomenological Landau theory for the helical Fermi liquid on the surface of a 3D topological insulator~\cite{PhysRevLett.115.066401}. In this context the dimensionless ``projected'' Landau parameters $\bar{F}_l$ are defined as $\bar{F}_l=\frac{1}{2}\mathcal{N}(\mu)f_l$ for $l\geq 1$, where $f_l$ is the quasiparticle interaction strength in angular momentum channel $l$. The difference in sign arises simply from the fact that in Eq.~(\ref{Hint}) an attractive interaction corresponds to $f_2>0$, while in Ref.~\cite{PhysRevLett.115.066401} it corresponds to $f_2<0$.

\begin{figure}
\includegraphics[width=1.0\linewidth]{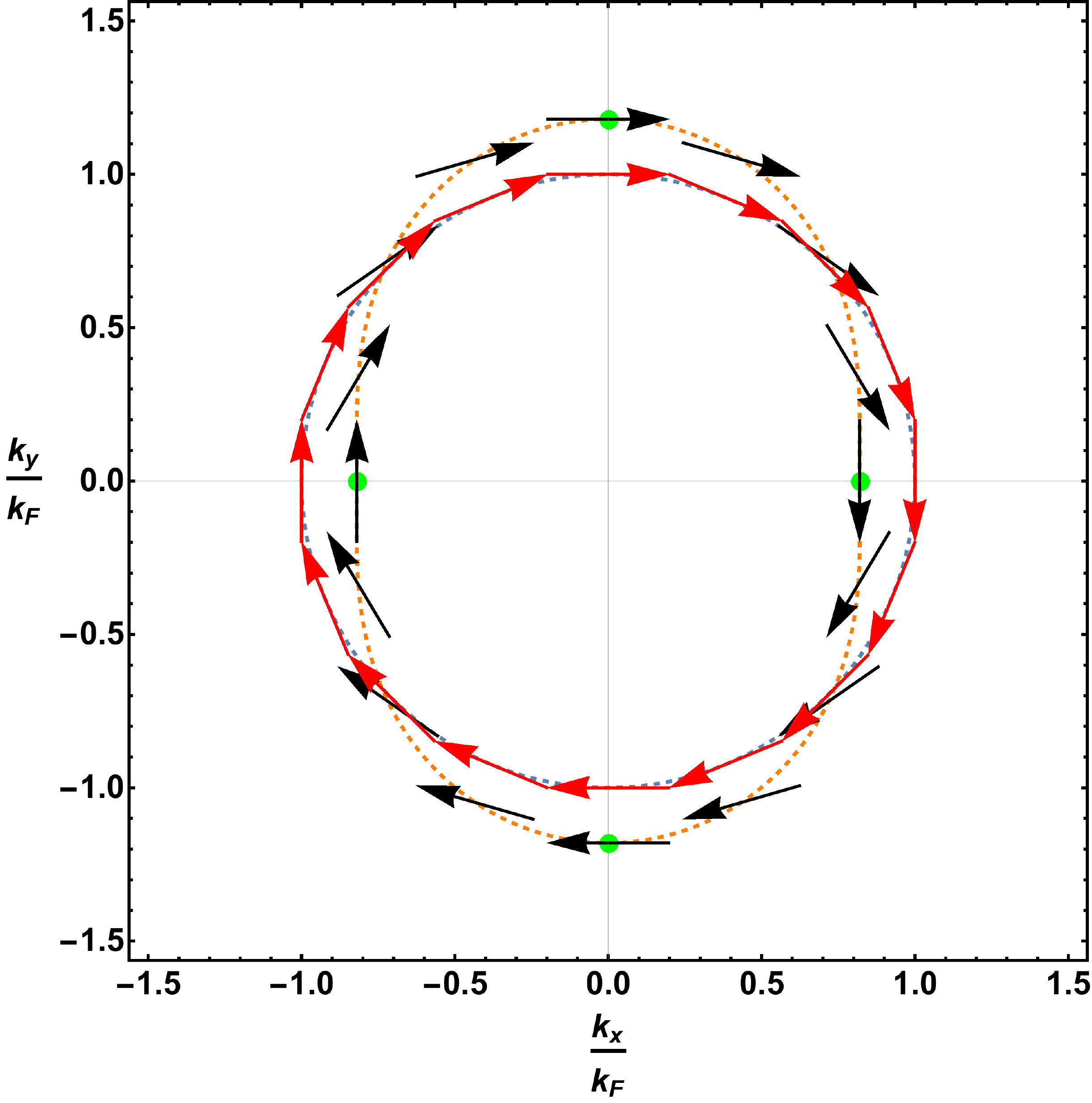}
\caption{(color online) Partial breakdown of spin-momentum locking in the nematic phase. Blue dashed line: Fermi surface in the isotropic phase ($\Delta_F=0$); orange dashed line: Fermi surface in the nematic phase (here shown for $\Delta_F=0.18$). The red (black) vectors represent the expectation value of spin on the Fermi surface in the isotropic (nematic) phase. Except at four special momenta (green dots), spin in the nematic phase is no longer perpendicular to momentum, nor is it tangential to the (distorted) Fermi surface.}
\label{vector_field}
\end{figure}

A first observable signature of nematic order of the type we have described in the doped limit is the partial breakdown of spin-momentum locking. In the doped limit, the mean-field Hamiltonian for fermionic quasiparticles is $H_\text{MF}=\sum_{\b{k}}\psi^{\dagger}_{\b{k}}\c{H}_\b{k}^{\phantom{\dagger}}\psi_{\b{k}}^{\phantom{\dagger}}$ where
\begin{align}
\c{H}_\b{k}=v_F\hat{\b{z}}\cdot(\b{\sigma}\times\b{k})-\mu+\bar{Q}_{ab}(\sigma_a\hat{k}_b+\sigma_b\hat{k}_a-\delta_{ab}\b{\sigma}\cdot\hat{\b{k}}),
\end{align}
and $\hat{k}_a=k_a/k$. Without loss of generality we choose $\bar{Q}_{12}=\bar{Q}_{21}=\bar{Q}$, $\bar{Q}_{11}=-\bar{Q}_{22}=0$, and thus
\begin{align}
\c{H}_\b{k}=v_F\hat{\b{z}}\cdot(\b{\sigma}\times\b{k})-\mu+2\bar{Q}(\sigma_x \hat{k}_y+\sigma_y \hat{k}_x).
\label{mean_electron}
\end{align}
Eq.~(\ref{mean_electron}) describes an anisotropic Fermi surface. Near the Fermi surface, the eigenstates have positive helicity (assuming $\mu>0$, thus above the Dirac point) and are given by
\begin{align}
|\psi_+(\b{k})\rangle=\frac{1}{\sqrt{2}}\left(
\begin{array}{cc}
ie^{i\theta_{\b{k}}}\frac{f(\theta_{\b{k}},\Delta_F)}{e^{2i\theta_{\b{k}}}-\Delta_F} \\ 
1
\end{array}\right),
\label{eigenstate}
\end{align}
where we define
\begin{align}
f(\theta_{\b{k}},\Delta_F)\equiv\sqrt{1+\Delta_F^2-2\Delta_F\cos 2\theta_{\b{k}}}.
\end{align}
We introduce a new dimensionless order parameter $\Delta_F\equiv 2\bar{Q}/\mu$ for the doped limit. The expectation value $\b{s}_\b{k}\equiv\langle\psi_+(\b{k})|\bsigma|\psi_+(\b{k})\rangle$ of the spin operator on the Fermi surface is in plane, with components
\begin{align}
s_{\b{k}}^x=\frac{(1+\Delta_F)\sin\theta_\b{k}}{f(\theta_{\b{k}},\Delta_F)}, \hspace{5mm}
s_{\b{k}}^y=-\frac{(1-\Delta_F)\cos\theta_\b{k}}{f(\theta_{\b{k}},\Delta_F)},\label{spin}
\end{align}
thus nematic order affects the spin polarization on the Fermi surface. To leading order in $\Delta_F$, the angle $\delta(\theta_\b{k})$ between the spin vectors in the presence and absence of nematic order is
\begin{align}
\delta(\theta_\b{k})\approx\Delta_F|\sin 2\theta_\b{k}|.
\end{align}
Thus except for four points on the Fermi surface $\theta_\b{k}=0,\pi/2,\pi,3\pi/2$, spin and momentum are no longer orthogonal (Fig.~\ref{vector_field}). However, one might naively think that spin-momentum locking is preserved in the sense that the spin vector remains tangent to the Fermi surface even if the latter is distorted. This is not true: defining a unit vector $\hat{\b{t}}_\b{k}$ tangent to the distorted Fermi surface (that winds around the Fermi surface clockwise), we have
\begin{align}
\hat{\b{z}}\cdot(\b{s}_\b{k}\times\hat{\b{t}}_\b{k})\approx\Delta_F\sin 2\theta_\b{k},
\end{align}
to leading order in $\Delta_F$, thus the spin vector is tangent to the Fermi surface only at four points, $\theta_\b{k}=0,\pi/2,\pi,3\pi/2$ (Fig.~\ref{vector_field}). This partial breakdown of spin-momentum locking except at high-symmetry points could be detected experimentally using spin-resolved angle-resolved photoemission spectroscopy (ARPES), using for instance the setups described in Ref.~\cite{xia2009,*hsieh2009,*pan2011}.

As in the undoped case, nematic order of the type considered here would lead to anisotropy in the in-plane spin susceptibility. Here the transition is continuous already at zero temperature, and in the vicinity of the zero temperature quantum critical point we find
\begin{align}
\chi_{xx}-\chi_{yy}=\frac{1}{4}g^2\mu_B^2\mathcal{N}(\mu)\frac{\Lambda}{k_F}\Delta_F,
\end{align}
to leading order in $\Delta_F$. More conventional measures of nematicity, such as anisotropy in the in-plane resistivity~\cite{oganesyan2001,PhysRevB.68.245109}, apply here as well. Considering scattering on nonmagnetic impurities modelled by a collision time $\tau$, a calculation of the conductivity using the Kubo formula and impurity-averaged Green's functions in the first Born approximation gives
\begin{align}
\frac{\rho_{xx}-\rho_{yy}}{\rho_{xx}+\rho_{yy}}\approx\Delta_F,
\end{align}
to leading order in $\Delta_F$ and assuming weak disorder $1/(\mu\tau)\ll 1$. By symmetry we anticipate an analogous result in the undoped case.

\section{Fluctuation effects}\label{sec:fluctuations}

We now go beyond the mean-field level and investigate the effect of fluctuations in the vicinity of the quantum critical point in the doped limit $k_F\gg\Lambda$. Following Ref.~\cite{oganesyan2001}, we rewrite the order parameter in terms of the Pauli matrices $\tau_z$ and $\tau_x$,
\begin{equation}
\hat{Q}=\psi^\dag\Delta_1\psi\tau_z+\psi^\dag\Delta_2\psi\tau_x,
\end{equation}
where 
\begin{equation}\label{Delta1and2}
\Delta_1=-i(\sigma_x\hat{\partial}_x-\sigma_y\hat{\partial}_y),\hspace{5mm}
\Delta_2=-i(\sigma_x\hat{\partial}_y+\sigma_y\hat{\partial}_x),
\end{equation}
and we define $\hat{\b{\partial}}\equiv\overset{\leftrightarrow}{\b{\partial}}/|\b{\partial}|$ in the sense of Fourier transforms (see Eq.~(\ref{OP})). We can now rewrite the imaginary-time action in a vectorial form,
\begin{equation}
S[\psi^\dag,\psi]=\int_0^{1/T}d\tau\int d^2r\left[\psi^\dag\hat{\mathcal{G}}^{-1}_0\psi-\frac{f_2}{2}(\psi^\dag\b{\Delta}\psi)^2\right],
\end{equation}
where $\b{\Delta}=(\Delta_1,\Delta_2)$ and
\begin{equation}
\hat{\mathcal{G}}^{-1}_0=\partial_\tau-iv_F\hat{\b{z}}\cdot(\b{\sigma}\times\b{\partial})-\mu,
\end{equation}
is the noninteracting Green's operator. Introducing a bosonic auxiliary field $\b{n}=(n_1,n_2)$ to decouple the four-fermion term, we have
\begin{align}
&S[\psi^\dag,\psi,\b{n}]=\nonumber \\
&\hspace{5mm}\int_0^{1/T}d\tau\int d^2r\left[\psi^\dag(\hat{\mathcal{G}}^{-1}_0-\b{n}\cdot\b{\Delta})\psi+\frac{1}{2f_2}\b{n}^2\right].
\label{vector_action_full}
\end{align}
After integrating out the fermions to second order in $\b{n}$, we find the effective action
\begin{align}
S_{\mathrm{eff}}[\b{n}]=
\frac{1}{2}\sum_{iq_n,\b{q}}\b{n}(\b{q},iq_n)^T\chi^{-1}(\b{q},iq_n)\b{n}(-\b{q},-iq_n),
\label{vector_action}
\end{align}
where the inverse propagator for the auxiliary field is given to lowest order in momentum $\b{q}$ and Matsubara frequency $q_n$ by
\begin{equation}\label{InversePropagator}
\chi^{-1}_{ij}(\b{q},iq_n)=\delta_{ij}(r+\kappa q^2)+M_{ij}(\b{q},iq_n).
\end{equation}
Here $r=f_2^{-1}-\mathcal{N}(\mu)/2$ is the distance from criticality which gives a mass to the auxiliary field, $\kappa=\mathcal{N}(\mu)/(8k_F^2)$ gives it a finite velocity, and 
\begin{align}
&M(\b{q},iq_n)=is\mathcal{N}(\mu)\int_0^{2\pi}\frac{d\phi}{2\pi}\frac{1}{is-\cos(\phi-\theta_{\b{q}})}\nonumber \\
&\times\bigg(\begin{array}{cc}
\sin^22\phi & -\sin 2\phi\cos 2\phi \\
-\sin 2\phi\cos 2\phi & \cos^22\phi
\end{array}\bigg),
\end{align}
is a dynamical term
where $s\equiv q_n/(v_Fq)$ and $\theta_{\b{q}}$ is the angle between $\b{q}$ and the $x$ axis. Performing the integral over $\phi$, we have
\begin{align}
&M(\b{q},iq_n)=\frac{\mathcal{N}(\mu)}{2}\frac{|s|}{\sqrt{s^2+1}}\nonumber\\
&\times\left[1-\left(\sqrt{s^2+1}-|s|\right)^4\left(\sigma_z\cos4\theta_\b{q}+\sigma_x\sin4\theta_\b{q}\right)\right],
\end{align}
which, after a rotation of $\theta_\b{q}$ by $\pi/4$, gives the same inverse propagator as for the spinless nematic Fermi fluid~\cite{oganesyan2001}. The effective action (\ref{vector_action}) can be diagonalized by a rotation $\b{n}\rightarrow\b{n}'$, $\chi^{-1}\rightarrow\chi^{\prime -1}$, where
\begin{align}
\b{n}'(\b{q},iq_n)&=R(4\theta_\b{q})^T\b{n}(\b{q},iq_n)\nn\\
&=\left(\begin{array}{c}
\hat{\b{d}}_\b{q}\cdot\b{n}(\b{q},iq_n) \\
\hat{\b{z}}\cdot\left(\hat{\b{d}}_\b{q}\times\b{n}(\b{q},iq_n)\right)
\end{array}\right).
\end{align}
Here $R(\phi)=e^{-i\sigma_y\phi/2}$ is an orthogonal rotation matrix and $\hat{\b{d}}_\b{q}\equiv(\cos2\theta_\b{q},\sin2\theta_\b{q})$. Thus $n_1'$ and $n_2'$ correspond to the longitudinal and transverse components of $\b{n}$, respectively. The transformed inverse propagator is
\begin{align}
\chi^{\prime -1}(\b{q},iq_n)&=R(4\theta_\b{q})^T\chi^{-1}(\b{q},iq_n)R(4\theta_\b{q})\nn\\
&=\left(\begin{array}{cc}
\chi^{\prime -1}_1(\b{q},iq_n) & 0 \\
0 & \chi^{\prime -1}_2(\b{q},iq_n)
\end{array}\right).
\end{align}
For small $s$, we have
\begin{align}
\chi^{\prime -1}_1(\b{q},iq_n)&=r+\kappa q^2+2\mathcal{N}(\mu)s^2+\ldots,\label{ChiInv1}\\
\chi^{\prime -1}_2(\b{q},iq_n)&=r+\kappa q^2+\mathcal{N}(\mu)|s|+\ldots\label{ChiInv2}
\end{align}

\subsection{Collective modes}

Since the inverse propagator of nematic fluctuations is the same as in the spinless case, the number and dispersion of collective modes, given by the condition
\begin{equation}
\det\chi^{-1}(\b{q},iq_n)=0,
\end{equation}
is also the same. Analytically continuing Eq.~(\ref{ChiInv1})-(\ref{ChiInv2}) to real frequencies $iq_n\rightarrow\omega+i\delta$, we find
\begin{align}
\chi_1^{\prime-1}(\b{q},\omega)&=r+\kappa q^2-2\mathcal{N}(\mu)\bigg(\frac{\omega}{v_Fq}\bigg)^2, \\
\chi_2^{\prime-1}(\b{q},\omega)&=r+\kappa q^2-\mathcal{N}(\mu)\frac{i\omega}{v_Fq},
\label{modes_crit_real}
\end{align} 
to leading order in $\omega/(v_Fq)$. At criticality $r\rightarrow 0^+$, the collective mode dispersions are
\begin{align}\label{CollectiveModeDisp}
\omega_1(q)\approx\sqrt{\frac{\kappa}{2\mathcal{N}(\mu)}}v_Fq^2,\hspace{5mm}
\omega_2(q)\approx-\frac{iv_F\kappa}{\mathcal{N}(\mu)}q^3,
\end{align}
thus $\omega_1$ is an undamped $z=2$ mode and $\omega_2$ is an overdamped $z=3$ mode. Since $\omega_2\ll\omega_1$ in the long-wavelength limit $q\rightarrow 0$, the overdamped mode dominates the long-wavelength response and the dynamical critical exponent at the transition is $z=3$~\cite{oganesyan2001}.

We note that although $\omega_1$ corresponds to longitudinal fluctuations of $\b{n}$, when projecting to the Fermi surface the longitudinal (11 and 22) components of the order parameter (\ref{OP}) map to the transverse (12 and 21) components of the usual spinless nematic order parameter
\begin{align}\label{spinlessOP}
\psi^\dag(\partial_a\partial_b-\frac{1}{2}\delta_{ab}\partial^2)\psi,
\end{align}
where the effectively spinless field $\psi^\dag$ creates electrons of the appropriate helicity on the Fermi surface (see Sec. S4 of the Supplemental Material in Ref.~\cite{PhysRevLett.115.066401}). Likewise, under projection the transverse components of (\ref{OP}) are mapped to the longitudinal components of (\ref{spinlessOP}). Thus in this sense $\omega_1$ ($\omega_2$) is the transverse (longitudinal) mode, in accordance with the terminology of Ref.~\cite{oganesyan2001}.

In the nematic phase ($r<0$), we consider Gaussian fluctuations about the classical saddle point, which we take to be $\bar{\b{n}}=(\bar{n},0)$ without loss of generality. Near the critical point where $\bar{n}$ is small, the leading change in the effective action for fluctuations compared to the isotropic phase is to the uniform and static part ($\b{q}=iq_n=0$) of the inverse propagator~\cite{oganesyan2001},
\begin{align}
&\chi^{-1}(\b{q},iq_n) \nonumber \\
&=\bigg(\begin{array}{cc}
 2|r|+\kappa q^2+M_{11}(\b{q},iq_n) & M_{12}(\b{q},iq_n)\\
M_{21}(\b{q},iq_n)& \kappa q^2+M_{22}(\b{q},iq_n)\end{array}\bigg),
\label{nem_prop}
\end{align}
i.e., the longitudinal (amplitude) mode $\delta n_1$ acquires a mass $2|r|$ and the transverse (Goldstone) mode $\delta n_2$ is massless. Deep in the nematic phase (i.e., $\bar{n}$ not small), the $q^2$ part and the dynamical part $M_{ij}$ will be modified from their form at $\bar{n}=0$, but our conclusions drawn from the small $\bar{n}$ limit will not be affected in a major way (for instance, a finite $\bar{n}$ would lead to a difference $\kappa_\perp\neq\kappa_\parallel$ in stiffness for the amplitude and Goldstone modes). The two eigenvalues $\chi^{-1}_\perp$ and $\chi^{-1}_\parallel$ of the inverse propagator (\ref{nem_prop}) give the spectrum of collective modes in the nematic phase. The inverse transverse propagator, given by
\begin{align}\label{GoldstonePropag}
&\mathcal{\chi}_{\perp}^{-1}(\b{q},iq_n)=\kappa q^2+\mathcal{N}(\mu)|s|\cos^2 2\theta_{\b{q}} \nonumber \\
&\hspace{5mm}-\mathcal{N}(\mu)\bigg(\cos 4\theta_{\b{q}}+\frac{\mathcal{N}(\mu)}{16|r|}\sin^2 4\theta_{\b{q}}\bigg)2s^2+\mathcal{O}(s^3),
\end{align}
corresponds to the gapless nematic Goldstone mode, which is overdamped due to Landau damping except along the principal axes of the distorted Fermi surface ($\theta_\b{q}=\pm\pi/4,\pm 3\pi/4$ for the saddle point considered, corresponding to $\bar{Q}_{11}=-\bar{Q}_{22}\neq 0$). Along those directions the inverse transverse propagator reduces to Eq.~(\ref{ChiInv1}) and the Goldstone mode disperses quadratically according to $\omega_1(q)$ in Eq.~(\ref{CollectiveModeDisp}). Those undamped directions also correspond to the Fermi surface momenta where spin-momentum locking is preserved (green dots in Fig.~\ref{vector_field}). The inverse longitudinal propagator is given by
\begin{align}\label{LongitudinalInvPropag}
&\mathcal{\chi}_{||}^{-1}(\b{q},iq_n)=2|r|+\kappa q^2+\mathcal{N}(\mu)|s|\sin^2 2\theta_{\b{q}} \nonumber \\
&\hspace{5mm}+\mathcal{N}(\mu)\bigg(\cos 4\theta_{\b{q}}+\frac{\mathcal{N}(\mu)}{16|r|}\sin^2 4\theta_{\b{q}}\bigg)2s^2+\mathcal{O}(s^3),
\end{align}
and describes gapped amplitude fluctuations, as expected.

Despite the number and dispersion of collective modes being formally the same as in the spinless nematic Fermi fluid, their physical nature is very different: in the latter case only charge degrees of freedom fluctuate, while fluctuations of the spin-orbit-coupled nematic order parameter (\ref{OP}) strongly mix charge and spin. An important observable consequence of this difference is that nematic fluctuations in the helical liquid considered here should strongly couple to the spin sector. While static nematic order does not break time-reversal symmetry and thus cannot induce a static spin polarization, nematic fluctuations can in principle induce spin fluctuations. To quantify this effect, one can use linear response: a nematic fluctuation $\delta\b{n}(\b{q},\omega)$ with momentum $\b{q}$ and frequency $\omega$ should induce a spin fluctuation $\delta\langle\b{s}(\b{q},\omega)\rangle$ with the same momentum and frequency,
\begin{align}\label{SpinFluct}
\delta\langle s_i(\b{q},\omega)\rangle\propto\Pi_{ij}^R(\b{q},\omega)\delta n_j(\b{q},\omega),
\end{align}
if a suitably defined retarded spin-nematic susceptibility $\Pi_{ij}^R(\b{q},\omega)$ is nonzero. An appropriate definition is
\begin{align}\label{SpinNematicChi}
\Pi_{ij}^R(\b{r},t)=-i\theta(t)\left\langle\left[(\psi^\dag\sigma_i\psi)_{(\b{r},t)},(\psi^\dag\Delta_j\psi)_{(\b{0},0)}\right]\right\rangle,
\end{align}
in real space and time, where $\psi^\dag\b{\sigma}\psi$ is the spin operator and $\psi^\dag\b{\Delta}\psi$ is the operator that couples to nematic fluctuations in Eq.~(\ref{vector_action_full}). Eq.~(\ref{SpinNematicChi}) will differ in the isotropic and nematic phases; here we compute $\Pi_{ij}^R$ in the isotropic phase and find a nonzero result, but we expect a nonzero result in the nematic phase as well. 

\begin{figure}
\includegraphics[width=1.0\linewidth]{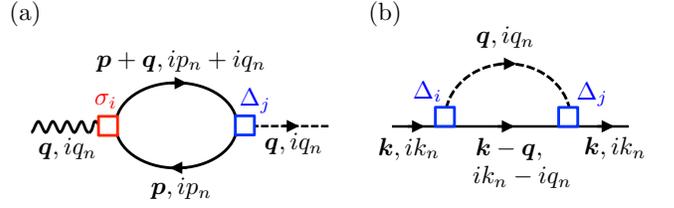}
\caption{One-loop diagrams for (a) the spin-nematic susceptibility [Eq.~(\ref{chiMatsubara})]; (b) the electron self-energy [Eq.~(\ref{FSE})].}
\label{fig:diagrams}
\end{figure}

In the Matsubara frequency domain, the spin-nematic susceptibility is given by the bubble diagram in Fig.~\ref{fig:diagrams}(a),
\begin{align}\label{chiMatsubara}
\Pi_{ij}(\b{q},iq_n)&=\frac{T}{V}\sum_{\b{p},ip_n}\tr\sigma_i\c{G}_0(\b{p}+\b{q},ip_n+iq_n)\nn\\
&\phantom{=}\times\Delta_j(\b{p},\b{p}+\b{q})\c{G}_0(\b{p},ip_n),
\end{align}
where
\begin{align}
\Delta_1(\b{k},\b{k}')&\equiv\sigma_x\left(\frac{\hat{k}_x+\hat{k}'_x}{2}\right)-\sigma_y\left(\frac{\hat{k}_y+\hat{k}'_y}{2}\right),\\
\Delta_2(\b{k},\b{k}')&\equiv\sigma_x\left(\frac{\hat{k}_y+\hat{k}'_y}{2}\right)+\sigma_y\left(\frac{\hat{k}_x+\hat{k}'_x}{2}\right),
\end{align}
are the Fourier transform of the nematic vertices (\ref{Delta1and2}) and $\c{G}_0$ is the unperturbed electron Green's function, given by
\begin{align}
\c{G}_0(\b{p},ip_n)=\frac{ip_n+\mu+v_F\hat{\b{z}}\cdot(\bsigma\times\b{p})}{(ip_n+\mu)^2-v_F^2\b{p}^2}.
\end{align}
The retarded spin-nematic susceptibility $\Pi_{ij}^R(\b{q},\omega)$ is obtained from (\ref{chiMatsubara}) by analytic continuation $iq_n\rightarrow\omega+i\delta$. We evaluate its imaginary part $\Pi_{ij}''(\b{q},\omega)$ at zero temperature and in the long-wavelength $q\ll k_F$, low-energy $|\omega|\ll\mu$ limits. To leading order in $\omega/v_Fq$, we find
\begin{align}
\Pi_{11}''(\b{q},\omega)&=-\Pi_{22}''(\b{q},\omega)\sim\frac{\omega}{v_F^2}\cos 3\theta_\b{q},\label{chi11}\\
\Pi_{12}''(\b{q},\omega)&=\Pi_{21}''(\b{q},\omega)\sim\frac{\omega}{v_F^2}\sin 3\theta_\b{q},\label{chi12}
\end{align}
ignoring constant prefactors (we are only interested in showing that the response does not vanish).
From time-reversal symmetry one can show that $\Pi_{ij}''(\b{q},\omega)=\Pi_{ij}''(-\b{q},-\omega)$, which is obeyed since Eq.~(\ref{chi11})-(\ref{chi12}) are odd in both $\b{q}$ and $\omega$. Kramers-Kronig relations imply that the real part $\Pi_{ij}'(\b{q},\omega)$ approaches a constant at low frequencies and has the same structure in momentum space. By virtue of Eq.~(\ref{SpinFluct}), nematic fluctuations can thus induce spin fluctuations, by contrast with the spinless (or spin degenerate) nematic Fermi fluid.

\subsection{Helical non-Fermi liquid behavior}

We now turn to the fermion self-energy on the Fermi surface. In the random phase approximation (RPA), i.e., at the one-loop level, the self-energy is given by the diagram in Fig.~\ref{fig:diagrams}(b),
\begin{align}
\Sigma(\b{k},ik_n)&=\frac{T}{V}\sum_{\b{q},iq_n}\sum_{ij}\Delta_i(\b{k},\b{k}-\b{q})\mathcal{G}_0(\b{k}-\b{q},ik_n-iq_n)\nn\\
&\phantom{=}\times\Delta_j(\b{k}-\b{q},\b{k})\chi_{ij}(\b{q},iq_n),
\label{FSE}
\end{align}
where $\chi_{ij}$ is the propagator of nematic fluctuations given in Eq.~(\ref{InversePropagator}). Here we only consider the effect of longitudinal fluctuations (i.e., the $z=3$ overdamped mode) which are expected to dominate at low energies. At the critical point $r=0$, we find
\begin{align}\label{SelfQCP1}
\Sigma(\b{k},ik_n)=\left(1+\hat{\b{z}}\cdot(\bsigma\times\hat{\b{k}})\right)\Sigma_0(\b{k},ik_n),
\end{align}
for $|k-k_F|\ll k_F$ and $|k_n|\ll\mu$, where
\begin{align}\label{SelfQCP2}
\Sigma_0(\b{k},ik_n)=-i\omega_0^{1/3}|k_n|^{2/3}\sgn k_n,
\end{align}
and $\omega_0\sim\mathcal{N}(\mu)^{-1}(v_F\kappa)^{-2}$, ignoring factors of order one. Near the Fermi surface, we can ignore the lower helicity branch (assuming $\mu>0$) and the electron Green's function $\c{G}(\b{k},ik_n)=\left[\c{G}_0(\b{k},ik_n)^{-1}-\Sigma(\b{k},ik_n)\right]^{-1}$ is given approximately by
\begin{align}
\c{G}(\b{k},ik_n)\approx\frac{1}{2}\frac{1+\hat{\b{z}}\cdot(\bsigma\times\hat{\b{k}})}{2i\omega_0^{1/3}|k_n|^{2/3}\sgn k_n-\xi_\b{k}},
\end{align}
where $\xi_\b{k}=v_F|\b{k}|-\mu$. Thus to a first approximation the critical Green's function retains the same helicity structure as in the noninteracting limit,
\begin{align}
\c{G}_0(\b{k},ik_n)\approx\frac{1}{2}\frac{1+\hat{\b{z}}\cdot(\bsigma\times\hat{\b{k}})}{ik_n-\xi_\b{k}},
\end{align}
but exhibits non-Fermi liquid behavior with vanishing quasiparticle residue as $\omega\rightarrow 0$. The spectral function is of the form
\begin{align}\label{SpecFuncCrit}
A(\b{k},\omega)\sim\frac{1}{2}\left(1+\hat{\b{z}}\cdot(\bsigma\times\hat{\b{k}})\right)\frac{\omega_0^{1/3}|\omega|^{2/3}}{\xi_\b{k}^2},
\end{align}
in the limit $\omega_0^{1/2}|\omega|^{2/3}\ll|\xi_\b{k}|\ll\mu$. Apart from the helicity structure, this is fully analogous to the spinless case~\cite{oganesyan2001}. In analogy with Ref.~\cite{PhysRevB.81.235105}, we conjecture that the transverse ($z=2$) fluctuations will give a finite anomalous dimension $\eta$ to the electron propagator, replacing the denominator $\xi_\b{k}^2$ in Eq.~(\ref{SpecFuncCrit}) by $|\xi_\b{k}|^{2-\eta}$.

In the nematic phase, the longitudinal modes are gapped [see Eq.~(\ref{LongitudinalInvPropag})] and one must look at the effect of the transverse Goldstone modes described by the inverse propagator (\ref{GoldstonePropag}). Because the symmetry generator $J_z$ that is broken in the nematic phase does not commute with translations, on general grounds one expects non-Fermi liquid behavior in the nematic phase as well~\cite{2014PNAS..11116314W}. By contrast with the electron self-energy at the critical point (\ref{SelfQCP1})-(\ref{SelfQCP2}) however, we expect the self-energy in the nematic phase to reflect the broken rotational symmetry.

To estimate the self-energy in the nematic phase, we observe that on the Fermi surface $|\b{k}|=k_F$, the electron Green's function appearing in Eq.~(\ref{FSE}) can be approximated by
\begin{align}\label{Gkq}
\c{G}_0(\b{k}-\b{q},ik_n-iq_n)\approx\frac{1}{2}\frac{1+\hat{\b{z}}\cdot(\bsigma\times\hat{\b{k}})}{ik_n-iq_n+v_F\hat{\b{k}}\cdot\b{q}},
\end{align}
since the momentum $\b{q}$ of the collective mode is much smaller than the Fermi momentum. Here we assume we are close to the quantum critical point such that the distortion of the Fermi surface is small and can be neglected in the calculation of the self-energy; this is an $\c{O}(\bar{n})$ effect, and can be understood in mean-field theory (Sec.~\ref{sec:MFDoped}), whereas the breakdown of Fermi liquid theory in the nematic phase appears at ``zeroth'' order in $\bar{n}$ as will be seen.  In the low-energy limit (i.e., on the Fermi surface $k_n\rightarrow 0$) Eq.~(\ref{Gkq}) is peaked at $\theta_\b{q}=\theta_\b{k}\pm\pi/2$, thus in Eq.~(\ref{FSE}) one can replace $\theta_\b{q}$ in the Goldstone mode propagator (\ref{GoldstonePropag}) by $\theta_\b{k}\pm\pi/2$~\cite{PhysRevB.81.045110,PhysRevB.73.085101},
\begin{align}
\mathcal{\chi}_{\perp}^{-1}(\b{q},iq_n)\approx\kappa q^2+\mathcal{N}(\mu)|s|\cos^2 2\theta_{\b{k}}.
\end{align}
We obtain
\begin{align}\label{SelfNematic}
\Sigma(\b{k},ik_n)=&\left(1-\sigma_y\cos 3\theta_\b{k}-\sigma_x\sin 3\theta_\b{k}\right)|\cos 2\theta_\b{k}|^{-2/3}\nonumber\\
&\times\Sigma_0(\b{k},ik_n),
\end{align}
where $\Sigma_0$ is defined in Eq.~(\ref{SelfQCP2}). Ignoring the lower helicity branch, we obtain the Green's function
\begin{align}\label{GFNematicSelf}
\c{G}(\b{k},ik_n)\approx\frac{1}{2}\frac{1+\hat{\b{z}}\cdot(\bsigma\times\hat{\b{k}})}{2i\omega_0^{1/3}|\cos 2\theta_\b{k}|^{4/3}|k_n|^{2/3}\sgn k_n-\xi_\b{k}},
\end{align}
and the spectral function
\begin{align}\label{SpecFuncNematic}
A(\b{k},\omega)\sim\frac{1}{2}\left(1+\hat{\b{z}}\cdot(\bsigma\times\hat{\b{k}})\right)\frac{\omega_0^{1/3}|\cos 2\theta_\b{k}|^{4/3}|\omega|^{2/3}}{\xi_\b{k}^2},
\end{align}
which are analogous to the spinless results~\cite{oganesyan2001} apart from the helicity structure. Equations~(\ref{SelfNematic})-(\ref{SpecFuncNematic}) hold for generic angles $\theta_\b{k}\neq\pm\pi/4,\pm 3\pi/4$ on the Fermi surface away from the principal axes of the nematic. Along the principal axes $\theta_\b{k}=\pm\pi/4,\pm 3\pi/4$, we find that after projection to the upper helicity branch the self-energy scales as $\sim|\omega|^{3/2}$, as in Ref.~\cite{oganesyan2001}, corresponding to long-lived quasiparticles along those directions. Equations~(\ref{SpecFuncCrit}) and (\ref{SpecFuncNematic}) correspond to a ``helical non-Fermi liquid'' in which the destruction of long-lived quasiparticles over most (in the nematic phase) or all (at the quantum critical point) of the Fermi surface coexists with a Berry phase of $\pi$ in spin space.

\section{Conclusion}\label{sec:conclusions}

In this work, we have developed a field-theoretic description of nematic order for a single Dirac cone on the surface of a 3D topological insulator. Due to spin-orbit coupling present in topological insulators, the nematic order parameter for helical Fermi liquids involves both spin and momentum, in contrast to the case of regular Fermi liquids which just involves momentum. In the undoped limit at zero temperature, we found a first-order isotropic-nematic transition at the mean-field level, in contrast with the expectation of a continuous transition based on Landau theory. The transition becomes continuous at a finite-temperature tricritical point. In the doped limit the transition was found to be continuous even at zero temperature. The spin-orbit coupled nature of nematic order was shown to lead to the partial breakdown of spin-momentum locking on the distorted Fermi surface and anisotropy in the in-plane spin susceptibility in both the doped and undoped limits. The number and dispersion of collective modes in the doped limit, as well as the prediction of non-Fermi liquid behavior at the quantum critical point and in the nematic phase, were seen to be the same as for spin rotationally invariant nematic Fermi fluids. However, in the helical case it was shown that nematic fuctuations can induce spin fluctuations, owing once again to the spin-orbit coupled nature of nematic order in these systems.


\acknowledgements

R.L. thanks D. Lorshbough, L. Janssen, V. Chua, P. Nyugen, L. Savary and G. Fiete for fruitful discussions and is particularly indebted to M. Edalati for providing an introduction to nematic phases. R.L. is supported by National Science Foundation Graduate Fellowship award number 2012115499 and a NIST NRC Research Postdoctoral Associateship Award. J.M. is supported by NSERC grant number RGPIN-2014-4608, the CRC Program, CIFAR, and the University of Alberta.

\bibliography{diracnematic}

\end{document}